\documentclass[12pt]{article}

\usepackage{epsf}

\def\f{\frac}
\def\b{\beta}

\def\e{\varepsilon}

\def\be{\begin{equation}}
\def\ee{\end{equation}}
\def\bea{\begin{eqnarray}}
\def\eea{\end{eqnarray}}

\def\ba{\begin{array}}
\def\ea{\end{array}}

\def\l{\left}
\def\r{\right}
\def\Det{\mathrm{Det}}
\def\Tr{\mathrm{Tr}}
\def\Li{\mathrm{Li}}

\begin{document}
\begin{titlepage}
\begin{center}
{\Large \bf William I. Fine Theoretical Physics Institute \\
University of Minnesota \\}
\end{center}
\vspace{0.2in}
\begin{flushright}
FTPI-MINN-09/06 \\
UMN-TH-2736/09 \\
February 2009 \\
\end{flushright}
\vspace{0.3in}
\begin{center}
{\Large \bf Destruction of a metastable string by particle collisions
\\}
\vspace{0.2in}
{\bf A. Monin \\}
School of Physics and Astronomy, University of Minnesota, \\ Minneapolis, MN
55455, USA, \\
and \\
{\bf M.B. Voloshin  \\ }
William I. Fine Theoretical Physics Institute, University of
Minnesota,\\ Minneapolis, MN 55455, USA \\
and \\
Institute of Theoretical and Experimental Physics, Moscow, 117218, Russia
\\[0.2in]
\end{center}

\begin{abstract}
We calculate the probability of destruction of a metastable string by collisions
of the Goldstone bosons, corresponding to the transverse waves on the string.
We find a general formula that allows to determine the probability of the string
breakup by a collision of arbitrary number of the bosons. We find that the
destruction of a metastable string takes place only in collisions of even number
of the bosons, and we explicitly calculate the energy dependence of such process
in a two-particle collision for an arbitrary relation between the energy and the
largest infrared scale in the problem, the length of the critical gap in the
string.
\end{abstract}

\end{titlepage}

\section{Introduction}
Field configurations with the properties of a metastable relativistic string
arise in a number of models\,\cite{Vilenkin,Preskill,Shifman}. Such string can
be viewed as having two phases, 1 and 2, with their respective tensions $\e_1$
and $\e_2$, and the larger tension $\e_1$ corresponding to the `upper'
metastable phase 1. The transition to the lower phase with $\e_2 < \e_1$ is
inhibited by the energy (mass) $m$ associated with the interface between the two
phases. The transition from the upper to the lower phase proceeds, similarly to
the decay of  false vacuum\cite{vko,Coleman} through formation  in the initial
string 1 of a gap spanned by the phase 2 and subsequent expansion of the gap
occupied by the lower phase. Clearly, the classical expansion of such gap is
possible only starting from a finite critical length of the gap $\ell_c=2
m/(\e_1-\e_2)$ at which length the energy loss due to the mass of the two ends
of the gap is compensated by the gain due to the phase energy difference
$(\e_1-\e_2) \, \ell_c$. A situation, where the string completely breaks up,
such as e.g. a breakup of a QCD string with formation of a heavy quark-antiquark
pair at the ends, can be viewed as the $\e_2 \to 0$ limit of a two-phase string.

The nucleation of the critical gap is a tunneling process, which can be analyzed
in terms of  the low energy dynamics of such string which dynamics can be
described by the effective Nambu-Goto action
\be
S=\e_1 \, A_1 + \e_2 \, A_2 + m \, P,
\label{nga}
\ee
where $A_1$ ($A_2$) is the space-time area of the world sheet occupied by the
phase 1 (resp. 2), and $P$ is the space-time length of the interface between the
two phases. The action (\ref{nga}) applies at the length scales much larger than
any scale for the internal structure of the string or of the interface, which
typically  is the thickness of the string. Using this effective action the
calculation of the probability of the decay of the metastable phase is quite
similar to the treatment of false vacuum decay in the so-called thin wall
approximation\,\cite{vko,Coleman}, and its exponential power has been
calculated\cite{Vilenkin,Preskill,Shifman} as well as the pre-exponential
factor\,\cite{mv1}. The rate $\gamma_0$ of spontaneous nucleation of critical
gaps per unit length per unit time in a string living in $d$ space-time
dimensions is
given by
\be
\gamma_0=F^{d-2} \, {\e_1 - \e_2 \over 2 \, \pi} \exp \left (- \pi \, {m^2 \over
\e_1 - \e_2 } \right )~,
\label{g0}
\ee
where $F$ is a dimensionless function of the ratio $\e_2/\e_1$, smoothly varying
from $F=e/(2 \sqrt{\pi}) =0.7668\ldots$ at $\e_2 =0$ to $F \to 1$ at $\e_2 \to
\e_1$.

The excitations of the string, described by the action (\ref{nga}) are massless
Goldstone bosons corresponding to transverse waves on the string. One can
reasonably expect on general grounds that the presence of such waves in an
excited state of the metastable phase catalyzes its decay. This catalytic effect
of the Goldstone excitations gives rise to the recently calculated\,\cite{mv2}
thermal enhancement of the decay rate. It can be noted that the massless
Goldstone modes are present in a thermal state at arbitrary low temperature $T$,
so that the thermal effect expands in powers of $T$. In particular the leading
at low $T$ behavior of the catalysis factor $K$ in the thermal decay rate
$\gamma_T= K \, \gamma_0$ is given by
\be
K=1 + (d-2) \, {\pi^8 \over 450}\, \left ( {\e_1-\e_2 \over 3 \e_1-\e_2} \right
)^2 \, \l(\f{\ell_c T}{2}\r)^8 + O\left[(\ell_c T)^{12} \right ]~,
\label{klow}
\ee
and the expansion parameter is $\ell_c T$, i.e. the ratio of the size of the
critical gap to the typical thermal wavelength\footnote{The full
expression\,\cite{mv2} in fact diverges at $\ell_c T =1$, beyond which point the
treatment in Ref.\cite{mv2} is not applicable.}.

In this paper we consider the destruction of the metastable phase of the string
by collisions of the Goldstone bosons. We find that this problem in fact can be
solved using appropriate interpretation and extension of the thermal treatment.
Indeed, the enhancement of the decay rate, described by the factor $K$, is
nothing else than the contribution of the production of the critical gap in
collisions of the bosons that are present in the thermal bath. In particular, we
find
that the first temperature dependent term in the expansion (\ref{klow}) is
entirely due to the process of the critical gap creation in collision of two
particles in the limit, where their center of mass energy $E=\sqrt{s}$ is much
smaller than $1/\ell_c$. We further consider a modification of the approach of
Ref.\cite{mv2} by formally introducing a negative chemical potential for the
bosons, and considering the decay of such a thermal state. This allows to
separate the contribution of $n$-boson collisions with different $n$ in all
terms of the expansion in the temperature, and thus to find the dependence of
the critical gap production in those collisions for an arbitrary relation
between $\ell_c$ and the energy of the bosons. One of the results of our
consideration is that the string destruction takes place only in collisions of
even number of bosons and is absent at odd $n$. By performing a full calculation
we find an explicit expression for the (dimensionless) probability $W_2$ of the
critical gap creation in two-boson collisions at arbitrary values of the
parameter $E \, \ell_c$. Our final result for this process has especially simple
form for the decay of the string into `nothing' i.e. at $\e_2=0$:
\be
W_2=2\, \pi^2 \, \ell_c^2 \, \gamma_0 \, \left [ I_3 \left ( {E \ell_c \over 2}
\right ) \right ]^2~,
\label{w20}
\ee
with $I_3(x)$ being the standard modified Bessel function of the third order.

Although in the end we make no assumption about the value of $(E \ell_c)$, our
treatment in the present paper is limited by the condition
\be
E^2 \ll \e_1~,
\label{econd}
\ee
so that the exponential growth of the energy dependent factor in Eq.(\ref{w20})
(and in subsequent more general expressions) does not overcome the exponential
suppression in the factor $\gamma_0$, described by Eq.(\ref{g0}).

Unlike the thermal factor $K$, which is singular\,\cite{mv2} at $\ell_c T = 1$,
the two-boson production rate does not exhibit any singularity at any value of
the parameter $E \ell_c$, and we find similar smooth energy behavior for
$n$-boson processes as well. We thus conclude that the `explosion' of the
thermal rate at $\ell_c T = 1$ is a result of infinite number of processes
becoming important at this point, rather than due to a finite set of processes
with a limited range of $n$ developing large probability at the energy per
particle of the order of $1/\ell_c$.

The rest of the material in this paper is organized as follows. In the Section 2
we present a brief account of the Euclidean space-time calculation of the string
transition, and describe some general properties of the processes of string
destruction by particle collisions following from their relation to the
absorption in the forward scattering amplitudes. In the Section 3 we discuss the
interpretation of the thermal catalysis of the string transition as induced by
collisions of particles in the thermal state, and we derive from the first term
of the low-temperature expansion in Eq.(\ref{klow}) the low-energy limit of the
two-particle string destruction probability $W_2$. The Section 4 describes the
extension of the finite temperature analysis by introduction of a negative
chemical potential, which allows to identify the contribution to the catalysis
factor of the string destruction by individual $n$-particle collisions. The
derived general expression is then used in the Section 5 to find the
two-particle probability $W_2$ in all orders in the parameter $(E \ell_c)$.
Finally, the Section 6 contains a discussion and summary of the results of the
present paper.

\section{Euclidean formulation and general considerations for particle
collisions}
In complete analogy with the false vacuum decay\,\cite{Coleman,Callan,Stone},
the calculation of the string decay rate\,\cite{mv1,mv2} can be performed
entirely in terms of Euclidean path integration. Considering the string in the
metastable phase 1 of large length $X$ extended along the $x$ axis with the ends
fixed at $x= \pm X/2$ and  the transverse coordinates $z_i$ ($i=1,\ldots, d-2$)
fixed at zero at the ends of the string, $z_i(\pm X/2)=0$, one needs to
calculate the Euclidean path integral ${\cal Z}_b$ over the string variables
with the action (\ref{nga}) around the bounce configuration, i.e. a
configuration involving a patch of the phase 2. The decay probability in the
full space-time area of the world sheet is then given by the expression
\be
W_0 = 2 \, {{\rm Im} {\cal Z}_b \over {\cal Z}_0}
\label{zrat}
\ee
with ${\cal Z}_0$ being the path integral calculated around the `flat'
configuration of the string 1, i.e. over the configuration without the patch of
the lower phase.

For the spontaneous decay and the thermally enhanced decay (at $T \ell_c < 1$)
the patch has the shape of a disk with the radius
\be
R={\ell_c \over 2} = {m \over \e_1-\e_2}~.
\label{rc}
\ee
The action on the bounce, relative to the unperturbed string, is equal to $\pi
\, m^2/(\e_1-\e_2)$ and determines the exponential power for the spontaneous
decay rate in Eq.(\ref{g0}) and
also for all the processes discussed in the present paper.

The only difference between the calculation of the decay at finite temperature
$T$ and at $T=0$ arises at the level of calculating the pre-exponential factor
due to the functional determinant of the quadratic part of the action and
amounts to the standard treatment of the boundary conditions in (Euclidean) time
for the fluctuations: zero boundary conditions at large time for considering
zero temperature and periodic boundary conditions with period $\beta=1/T$ at
finite temperature.

The formula (\ref{zrat}) is in  fact the unitarity relation\,\cite{Stone}
between the decay rate and the imaginary part of the amplitude of the transition
amplitude from the false vacuum to the false vacuum $\langle ({\rm
vacuum~1})_{\rm out}\,| {\rm vacuum~1})_{\rm in} \rangle$. Similarly, one can
treat the probability of the string breakup by an excited state in terms of the
bounce contribution to the imaginary part of its forward scattering amplitude.
Proceeding to discussion of the string decay induced by the Goldstone bosons, we
readily notice that a state with just one Goldstone boson cannot induce the
destruction of the string. Indeed, the total probability of such induced process
is  Lorentz invariant and thus can depend only on the (Lorentz) square of the
particle momentum $k^2$. The Goldstone bosons are massless, so that for them
$k^2=0$ and is fixed. Thus if a single massless boson produced an effect on the
decay, this effect would have no dependence on the energy $\omega$ of the
boson. In the limit $\omega \to 0$ the Goldstone boson is indistinguishable from
the vacuum. (In other words, the limit $\omega \to 0$ corresponds to an overall
shift of the string in transverse direction.) Thus the decay rate of a single -
boson state is the same as that of the vacuum, and the presence of a single
boson with any energy produces no effect.

The simplest excited state, contributing to the string destruction, is that with
two particles. The probability $W_2$ of creation of the critical
gap\,\footnote{It should be emphasized that the considered probability is
inclusive in the sense that we make no assumption about any additional to the
critical gap products of the reaction.} in a  collision of two particles with
two-momenta $k_1$ and $k_2$ can depend only on the Lorentz invariant
$s=(k_1+k_2)^2$. Obviously, for two particles colliding on a string one has
$s=4 \, \omega_1 \omega_2$ (and $s=0$ for two particles moving in the same
direction,
i.e. non-colliding). Using the unitarity relation this probability can be found
in terms of the imaginary part of the forward scattering amplitude
$A(k_1,k_2;k_1,k_2)$:
\be
W_2 = C { {\rm Im} A(k_1,k_2;k_1,k_2) \over \omega_1 \omega_2}~
\label{w2u}
\ee
where the factor $\omega_1 \omega_2$ is the usual flux factor, and the constant
$C$ does not depend on either of the energies, and is determined by specific
convention about the normalization of the amplitude.

The dynamics of the Goldstone bosons on the string, including their scattering,
can be considered in terms of the transverse shift $z_i(x)$ treated as a
two-dimensional field and described by the Nambu-Goto action (\ref{nga}) as
\be
S=\e_1 \, \int_{A_1} \, \sqrt{1+ (\partial_\mu z_i)^2} \, d^2x +  \e_2 \,
\int_{A_2} \, \sqrt{1+ (\partial_\mu z_i)^2} \, d^2 x + m \int_P \sqrt{1+
(\partial z_i/\partial l)^2} \, dl~,
\label{ngg}
\ee
where $dl$ is the element of the length of the interface $P$ between the phases.

Clearly at low energy of the Goldstone bosons one can make use of the expansion
in Eq.(\ref{ngg}) in powers of $(\partial z)$ which generates the expansion of
the scattering amplitudes in the momenta of the particles with each one entering
the amplitude with (at least) one power of its energy $\omega$, as is mandatory
for Goldstone bosons. For the scattering in the metastable state this generates
expansion in powers of $\omega / \sqrt{\e_1}$, so that in the zeroth order in
this ratio, that we are discussing here, it is sufficient to retain only the
quadratic in $(\partial z)$ terms in the action (\ref{ngg}). It should be noted
that in spite of retaining only the quadratic terms, the multi-boson scattering
amplitudes do not vanish, since the necessary nonlinearity arises from the
bounce configuration. In other words, the bosons scatter `through the bounce'.
The energy expansion for these amplitudes is determined by the bounce scale
$\ell_c$, so that the parameter of such expansion is $(\omega  \ell_c)$, and we
do not assume this parameter to be small. Clearly,  the condition for
applicability of the
present approach where the terms of order $\omega / \sqrt{\e_1}$ are dropped,
while those with the parameter $\omega \ell_c$ are retained is that
\be
{m^2 \over \e_1-\e_2} \gg 1-{\e_2 \over \e_1}~,
\label{condr}
\ee
which is always true if the semiclassical tunneling can be applied at all to the
string decay.

We shall now show that in the on-shell scattering through the bounce each
external leg enters with at least two powers of its energy. Let us start, for
the simplicity of illustration, with the binary scattering. The general two
$\to$ two scattering amplitude $ A(k_1,k_2;k_3,k_4)$ can be related in the
standard application of the reduction formula to the connected 4-point Green's
function $\langle {\rm vacuum~1}| T\{z(x_1) z(x_2) z(x_3) z(x_4) \}| {\rm
vacuum~1} \rangle$, which in turn is an analytical continuation of the Euclidean
connected correlator
\be
\langle {\rm vacuum~1}| z(x_1) z(x_2) z(x_3) z(x_4) | {\rm vacuum~1} \rangle =
{\cal Z}_0^{-1} \, {\delta^4 \, {\cal Z}_b[j] \over \delta j(x_1)\, \delta
j(x_2) \, \delta j(x_3) \, \delta j(x_4) } \left . \right |_{j=0}~.
\label{zj}
\ee
The latter expression for the correlator implicitly assumes the conventional
procedure of introducing in the action the source term $\int j(x) \, z(x) \,
d^2x$ for the Goldstone variable $z(x)$\,\footnote{We temporarily suppress the
spatial index of the transverse shift variable $z_i(x)$, which is equivalent to
considering just one transverse dimension, i.e. $d=3$.}, and ${\cal Z}_b[j]$ is
the path integral around the bounce in the presence of the source.

\begin{figure}[ht]
  \begin{center}
    \leavevmode
    \epsfxsize=12cm
    \epsfbox{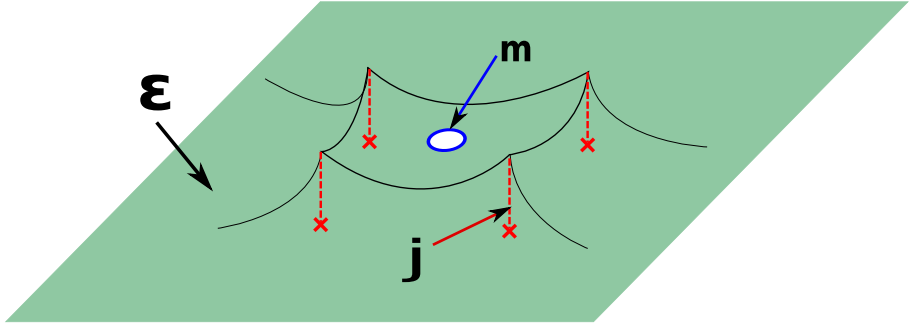}
    \caption{Bounce configuration distorted by the sources}
  \end{center}
\label{4_source}
\end{figure}

The low-energy limit of the on-shell scattering amplitude is determined by the
correlator at widely separated points $x_1,\ldots,x_4$. The weak $\delta$-
function sources `prop' the string in the transverse direction at those points
as shown in Figure~1, and generally distort the bounce located between the
sources. The correlator (\ref{zj}) is determined by the distortion of the bounce
by all four sources, so that at large separation between the points the bounce
is located far (in the scale of its size $\ell_c$) from the sources, where the
overall distortion of the world sheet for the string is small and is slowly
varying on the scale $\ell_c$. One can therefore expand the background field
$z_s$ generated by the sources at the bounce location in the Taylor series
around an arbitrarily chosen inside the bounce point $x=0$. Clearly the first
term of this expansion $z_s(0)$ corresponds to an overall transverse shift of
the string and does not alter the bounce shape and the action. Moreover, the
linear term in this expansion, proportional to the gradient $\partial z_s(0)$,
does not change the bounce action either. Indeed this term corresponds to a
linear incline of the string in the transverse coordinates, and can be
eliminated by an overall rotation of the string in the $d$ dimensional space. In
other words the absence of the linear in the gradient of $z_s$ term in the
action is a direct consequence of the $d$-dimensional Lorentz invariance of the
string. We thus arrive at the conclusion that the expansion for the distortion
of the bounce starts from the second order in the derivatives of $z_s$ (the
curvature of the background world sheet), which for a connected correlator
implies that in the expansion in the energy each external source enters with at
least two powers of the energy. This conclusion clearly applies to the on-shell
amplitudes with arbitrary number of external legs, since the generalization to
multiparticle scattering is straightforward.

In particular, the binary scattering amplitude $ A(k_1,k_2;k_3,k_4)$ at low
energy scales as the eighth power of the energy. It the follows from
Eq.(\ref{w2u}) that the probability of the string destruction by collision of
two
particles is proportional to the sixth power of the energy scale, or
equivalently, to $s^3$ at small $s$. Applying in the same manner the unitarity
condition to the forward scattering amplitude of a general $n$-particle state,
one readily concludes that the corresponding probability $W_n$ of the induced
string breakup scales with the overall energy scale $\omega$ as  $W_n \propto
\omega^{3n}$.

\section{Thermal decay and the string destruction by particle collisions}

The described procedure for calculating the bounce-induced scattering amplitudes
in terms of the Euclidean correlators runs into the technical difficulty of
calculating the bounce distortion in the background created by the sources.
Furthermore, this procedure obviously involves a great redundancy, if the final
purpose is a calculation of the probabilities $W_n$, i.e. only the absorptive
parts
of the forward scattering on-shell amplitudes are the quantities of interest. We
find that in fact one can directly derive the probabilities $W_n$ by an
appropriate interpretation of the more readily calculable thermal decay rate in
terms of the collision-induced probabilities. Such an approach is technically
more tractable due to the fact that in the thermal calculation the bounce is not
distorted, i.e. it is still a flat disk, and only the boundary conditions for
the modes of the fluctuations are modified. We first illustrate this approach by
using the first nontrivial term of the low temperature expansion in
Eq.(\ref{klow}) for a calculation of the low energy limit of the binary
probability $W_2$.

Indeed, the temperature dependent factors in the catalysis factor $K$ arise from
the critical gap nucleation in boson collisions weighed
with the thermal number density distribution for the massless Goldstone bosons
\be
d\,n(k) = {1 \over e^{\omega/T} - 1} \, {d \, k \over 2 \pi}~,
\label{bose}
\ee
where $k$ is the spatial momentum, $|k|=\omega$, running from $-\infty$ to
$+\infty$. Given the low energy behavior $W_n \propto \omega^{3n}$ found in the
previous section, one readily concludes that the contribution of $n$-particle
string destruction starts with the term $T^{4n}$ in the low $T$ expansion for
$K$. Thus the first term written in Eq.(\ref{klow}) can arise only from $n=2$,
i.e. from the production of the critical gap in binary collisions. Writing the
expansion in
$s$ for the probability $W_2$ as $W_2=c_3 \, s^3  + \ldots$, one determines the
coefficient $c_3$ of the $s^3$ term  by comparing the result in Eq.(\ref{klow})
with the one calculated in terms of the two-particle collision rate using the
number density distribution (\ref{bose}):
\bea
&&(d-2) \, \gamma_0 \, {\pi^8 \over 450}\, \left ( {\e_1-\e_2 \over 3 \e_1-\e_2}
\right
)^2 \, \l(\f{\ell_c T}{2}\r)^8 = \nonumber \\
&&(d-2) \, c_3 \, \int_0^\infty {d\omega_1 \over  e^{\omega_1/T} - 1} \,
\int_0^\infty {d \omega_2 \over  e^{\omega_2/T} - 1}\, {s^3 \over 4 \pi^2}
 = (d-2) \, c_3 \, {16 \, \pi^6 \over 225} \,  T^8~,
\label{w2low}
\eea
where the relation $s=4 \, \omega_1 \omega_2$ is used and the integration is
done
only for the bosons with opposite signs of their momenta, and avoiding the
double-counting for identical particles. The factor $(d-2)$ counting the
number of the transverse dimensions, corresponds to the summation over the
polarizations of the Goldstone bosons. The expression for the coefficient $c_3$
following from Eq.(\ref{w2low}) thus determines the first term in the expansion
for $W_2$
\be
W_2=\gamma_0 \, R^2 \, \left [ { \pi^2 \over 32}  \, \left ( {\e_1-\e_2 \over 3
\e_1-\e_2} \right
)^2 \, R^6 \, s^3 + \ldots \right ]~.
\label{w2l1}
\ee
(One can notice that at $\e_2=0$ the $s^3$ term coincides with the first term of
expansion of the expression in Eq.(\ref{w20}).)

\section{Thermal bath with a chemical potential}

The discussed procedure for extracting the coefficients of the energy expansion
for the probability of collision-induced string decay is obviously limited to
only the first term in $W_2$. In the higher terms in the temperature expansion
for $K$ the contribution of the energy expansion for $W_n$ with different $n$
generally gets entangled. This happens because the temperature is the only
parameter and terms originating from different $n$ can contribute in the same
power in $T$. In order to disentangle the terms of higher order in the energy in
$W_n$ with low $n$ from similar terms originating from higher $n$ we introduce a
{\em negative} chemical potential $\mu$ for the Goldstone bosons. Generally,
such procedure would no be possible, since the number of these bosons is not
conserved. However in our application this procedure is fully legitimate.
Indeed, the thermal state of the string that we study here is that of {\em
collisionless} bosons, in which their number {\em is} conserved. The string
decay, resulting in a change in this number, is a very weak process that we
consider only in the first order, which justifies averaging the rate of this
process over the unperturbed state with conserved number of particles. At
negative $\mu$ the number density distribution of the bosons (\ref{bose}) is
replaced by
\be
d \, n (k) = {1 \over e^{\omega + |\mu| \over T} - 1} \, {d \, k \over 2 \pi}~,
\label{bosem}
\ee
and by tuning the parameter $|\mu|/T$ one can readily resolve between the
contribution of $n$-particle processes with different $n$.

The introduction of the chemical potential requires us to modify our previous
thermal calculation\,\cite{mv2}. Fortunately, this modification is quite
straightforward, and we describe it here by first briefly recapitulating the
relevant part of that previous work, where further details can be readily found.

The bounce configuration is a disk of the radius $R$, and we place the origin of
the string world sheet coordinates in the Euclidean space-time $(x,t)$ at the
center of the bounce. A calculation of the pre-exponential factor in the path
integral ${\cal Z}_b$ requires finding the functional determinant of the second
variation of the action over the fluctuations of the Goldstone field $z_i(x)$.
The functional determinant for the fluctuations in each
transverse direction factorizes, so that it is sufficient to consider only one
Goldstone polarization $z(x)$. The eigenmodes of the second variation of the
action are harmonic functions of $(x,t)$, and they can thus be considered as the
real and imaginary parts of holomorphic functions of the complex variable
$w=t+i\,x$. For the problem at zero temperature the harmonics inside the bounce
(the `inner' ones) are generated in this way by the set of functions
\be
f_k^{\rm in}={w^k \over R^k}
\label{fkin}
\ee
with integer $k$ starting from $k=0$.
while those regular at
$|w| \to \infty$ (the `outer' ones) are given by the set of functions $R^k/w^k$.

At finite temperature $T$ the path integral runs over the configurations
periodic in (the Euclidean) time $t$ with the period $\beta=1/T$. At $T <
1/\ell_c$ the bounce fits within one period and the inner part of the
fluctuations is not affected by the thermal boundary conditions, while the outer
part can be taken as the set of functions corresponding to periodically summed
outer harmonics. For $k \ge 2$\,\footnote{The harmonics with $k=0$ and $k=1$ are
found\,\cite{mv2} to be inessential for the discussed thermal effects. This
property can in fact be traced to the discussed in Sect.2 insensitivity of the
bounce to a constant and linear shift of the background transverse coordinates
$z$.} these periodic  harmonics can be written as
\be
g_k(w)=\f{R^k}{w^k}+\sum_{n=1}^\infty \l [ \f{R^k}{(w-n\b)^k}+\f{R^k}{(w+n\b)^k}
\r ]~.
\label{gk}
\ee
The functions $g_k$ are expanded in powers of $w$ as
\be
g_k(w)=\f{R^k}{w^k}+\sum_{p=1}^\infty d_{pk}\l(\f{w}{R}\r)^p,
\label{g_exp}
\ee
with
\be
d_{pk}=\l [(-1)^k+(-1)^p\r]\l(\f{R}{\b}\r)^{k+p}\, {(p+k-1)! \over p! \, (k-1)!
}
\, \zeta(p+k)~,
\label{D_matrix}
\ee
where we use the standard notation $\zeta(q)=\sum_{n=1}^\infty n^{-q}$ for the
Riemann zeta function.

The matching of the outer eigenmodes with the inner ones given by
Eq.(\ref{fkin}) thus involves an additional (in comparison with the limit $T=0$)
term, described by the matrix $D$ with the elements $D_{pk} = d_{pk}$ for $p$
and $k$ starting from two. This is this additional contribution that produces
the thermal effect in the string decay rate, and the  catalysis factor $K$ can
be written in terms of the matrix ${\cal D}$ with the elements
\be
{\cal D}_{pk}= -\l[(-1)^k+(-1)^p\r] \l ( R T \r)^{k+p+2}\,{p \over p+b} \,
{(p+k+1)! \over (p+1)! \, k! } \, \zeta(p+k+2)~,
\label{cald}
\ee
where the indices $p$ and $k$ are shifted by one unit, so that their count
starts at one: $p,k=1,\ldots$, and the notation is used
$b=(\e_1+\e_2)/(\e_1-\e_2)$. The final expression expression\,\cite{mv2} for the
thermal factor $K$ in $d$ space-time dimensions reads as
\be
K=\Det\l[1-{\cal D}^2\r]^{-(d-2)/2}~.
\label{kfull}
\ee

The modification of this calculation for a thermal state with a negative
chemical potential, in which the number density of the bosons is given by
Eq.(\ref{bosem}), is achieved by introducing a `damping factor' in the periodic
sums for the outer functions in Eq.(\ref{gk}):
\be
g_k(w) \to g^{(\mu)}_k(w)
=\f{R^k}{w^k}+\sum_{n=1}^\infty\l[\f{R^k}{(w-n\b)^k}+\f{R^k}{(w+n\b)^k}\r] \,
\exp \left( - n \, |\mu| \, \beta \right ) ~.
\label{gkm}
\ee
One can readily see that the only net result of the so-introduced $\mu$
dependent factor in
the calculation of Ref.\cite{mv2} is a modification of the matrix coefficients
$d_{pk}$ amounting to the replacement of the factors $\zeta(q)$ by the
polylogarithm function,
$$\Li_q(x)=\sum_{n=1}^\infty { x^n \over n^q}~,$$
$\zeta(q) \to \Li_q(e^{-|\mu|/T})$. In other words, the catalysis factor for a
thermal state with a negative chemical potential is given by the expression
\be
K(\mu,T) = \Det\l[1-{\cal D}^2(\mu, T)\r]^{-(d-2)/2}~,
\label{kmt}
\ee
with the elements of the matrix ${\cal D}(\mu, T)$ having the form
\be
\left [ {\cal D}(\mu,T) \right ]_{pk}= -\l[(-1)^k+(-1)^p\r] \l ( R T
\r)^{k+p+2}\,{p \over p+b} \, {(p+k+1)! \over (p+1)! \, k! } \, \Li_{p+k+2}
\left ( e^{-|\mu|/T} \right )~.
\label{cdmt}
\ee

The dependence on two `tunable' parameters $\mu$ and $T$ in Eq.(\ref{kmt}) makes
it possible to disentangle the contribution of processes with different number
of particles from the energy behavior in each of these processes. Such a
separation becomes straightforward if one notices that each factor with the
polylogarithm $\Li$ arises from the integration over the distribution
(\ref{bosem}):
\be
\int_0^\infty {\omega^q \over e^{\omega + |\mu| \over T} - 1} \, d \omega = q!
\, T^{q+1} \, \Li_{q+1} \left ( e^{-|\mu|/T} \right )~.
\label{bosint}
\ee
One therefore concludes that the number of the `$\Li$ factors' in each term of
the expansion of the catalysis factor in Eq.(\ref{kmt}) directly gives the
number of particles in the process, while the indices of these polylogarithmic
factors give the power of the energy for each particle. Given that the matrix
${\cal D}(\mu,T)$ is linear in the `$\Li$ factors', one can count the number of
particles contributing to each term of the expansion for $K(\mu,T)$ by counting
the power of  ${\cal D}(\mu,T)$. The latter counting is simplified if
one rewrites the equation (\ref{kmt}) in the equivalent form, suitable for the
expansion in powers of ${\cal D}(\mu,T)$:
\bea
\label{dexp}
K(\mu,T) &=& \exp \left \{ - {d-2 \over 2} \, \Tr  \ln \left [1- {\cal
D}(\mu,T)^2 \right ] \right \} = \\  \nonumber
{d-2 \over 2} \, \Tr \left [ {\cal D}(\mu,T)^2 \right ] & + & {d-2 \over 4} \,
\Tr \left [ {\cal D}(\mu,T)^4 \right ] + {(d-2)^2 \over 8} \, \left \{ \Tr \left
[ {\cal D}(\mu,T)^2 \right ] \right \}^2 + O \left ( {\cal D}^6 \right ).
\eea

The latter expression merits some observations. The first is that all the terms
in the expansion in powers of ${\cal D}(\mu,T)$ are positive, which is certainly
the necessary condition for the consistency of our interpretation of these terms
as corresponding to the probability of the destruction of the string by $n$-
particle collisions. The second is that the string is destroyed only in
collisions of {\em even} number of particles, since the expansion manifestly 
goes in the even powers of ${\cal D}(\mu,T)$. Finally, the third observation is
related to the dependence in Eq.(\ref{dexp}) on the number of the transverse
dimensions $(d-2)$. Namely, the quadratic in ${\cal D}(\mu,T)$ term is
proportional to $(d-2)$. This corresponds to that in two-particle collisions
only the Goldstone bosons with the same transverse polarization do destroy the
string. On the contrary, the quartic in ${\cal D}(\mu,T)$ term has one
contribution proportional to $(d-2)$ and one proportional to $(d-2)^2$. The
linear in $(d-2)$ part  corresponds to  all the bosons in the collision having
the same polarization, while the quadratic in $(d-2)$ part is necessarily
contributed by the collisions, where the colliding bosons have different
polarizations.

\section{Destruction of the string in two-particle collisions at arbitrary $(s
R)^2$}

The expression (\ref{dexp}) for the catalysis factor $K(\mu,T)$ together with
the formulas (\ref{cdmt}) and (\ref{bosint}) reduce the calculation of the
probability of the string breakup by a collision of an arbitrary (even) number
$n$ of particles to straightforward, although not necessarily short, algebraic
manipulations. In this section we consider in full the most physically
transparent case of two-particle collisions. The probability in this case is
found from the term in Eq.(\ref{dexp}) with the trace $\Tr  [ {\cal D}(\mu,T)^2
]$. Using Eq.(\ref{cdmt}), this trace can be written as a double sum:
\bea
\label{trd2}
&&\Tr \left [ {\cal D}(\mu,T)^2 \right ] = \\
\nonumber
&&4 \, \sum_{p=1}^\infty \sum_{k=1}^\infty \, { 1 \over (p+b+1) \, (k+b+1) } \,
{\left[ (p+k+1)! \,(RT)^{p+k+2} \, \Li_{p+k+2} \left ( e^{-|\mu|/T} \right )
\right ]^2  \over (p-1)! \, (p+1)! \, (k-1)! \, (k+1)! }
\eea
One can readily recognize the expression in the straight braces here as the
integral (\ref{bosint}) with the power of the energy $q$ given by $(p+k+1)$, and
thus identify the coefficient of the same power of $s=4 \omega_1 \omega_2$ in
the expansion of the probability $W_2(s)$ in $s$. In this way we find the
following formula for $W_2(s)$ in terms of this expansion,
\bea
W_2(s) & = & 8 \, \pi^2 \, \gamma_0 \, R^2 \,  \sum_{p=1}^\infty
\sum_{k=1}^\infty \, { 1 \over (p+b+1) \, (k+b+1) } \, {(s R^2/4)^{p+k+1}  \over
(p-1)! \, (p+1)! \, (k-1)! \, (k+1)! } \nonumber \\
&=& 8 \, \pi^2 \, \gamma_0 \, R^2 \,  \left [ \Phi_b(\sqrt{s} \, R) \right ]^2~,
\label{w2fin}
\eea
where the function $\Phi_b(x)$ expands in a single series as
\be
\Phi_b (x) = { x \over 2} \sum_{p=1}^\infty {1 \over p+b+1} \, {x^{2p} \over
(p-1)! \, (p+1)! }~.
\label{phib}
\ee
It can be noted that the two-particle probability depends only on the odd powers
of $s$. This in fact is a consequence of the binary forward scattering amplitude
being even in $s$, as required by the Bose symmetry.

For integer values of the parameter $b$, the function $\Phi_b(x)$ has a simple
expression in terms of the modified Bessel functions $I_q(x)$ of the order $q$
up to $q=b+2$. This expression is especially simple for $b=1$, i.e. for the case
of the string decay into `nothing': $\Phi_1(x)=I_3(x)$, so that one arrives at
the formula (\ref{w20}). We also write here, for an illustration, the
corresponding expressions for the next two integer values of $b$:
\be
\Phi_2(x)={1 \over x} \, \left [ I_5(x) + 6 I_4(x) \right]\,;~~~~~~~
\Phi_3(x)={1 \over x^2} \, \left [ (48+ x^2) \, I_5(x) + x \, I_6(x) \right]\,~.
\label{phibex}
\ee

\section{Discussion and summary}

One can readily evaluate the sum in Eq.(\ref{phib}) at large $x$ by using the
Stirling formula for the factorials, and find that independently of $b$ the
function behaves as an exponent: $\Phi_b(x) \sim e^x$. This corresponds to an
exponential behavior of the probability $W_2$ at large $(s R^2)$. In combination
with the exponential factor in $\gamma_0$ this behavior can be written as
\be
W_2(s) \sim \exp \left ( -\, \pi \, {m^2 \over \e_1-\e_2} + 2 \sqrt{s} \, {m
\over \e_1-\e_2} \right )~.
\label{w2ex}
\ee
Clearly, the energy growth of the probability never overcomes the large negative
power due to the tunneling. Indeed, due to the condition $m^2 \gg \e_1$ for the
semiclassical approximation to be valid at all and the condition $s \ll \e_1$
for the validity of our approach based on neglecting direct interactions between
the Goldstone bosons, one always has $\sqrt{s} \ll m$.

It can be also noted that the power series for $\Phi_b(x)$ converges for any
value of $x$, so that the probability $W_2(s)$ has no singularity at any $s$.
The same conclusion also holds for the probabilities $W_n$ at any $n$. Indeed,
each term of order ${\cal D}^n$ with a specific power $n$ in the expansion
(\ref{dexp}) is described by an absolutely convergent series in the energy
parameters. This implies that the reason for the singularity\cite{mv2} at $T
\ell_c = 1$ of the thermal factor $K$ in Eq.(\ref{kfull}) is not that some
finite set of collision processes gives a singular contribution, but rather that
at this point an infinite multitude of processes becomes important.

In summary. We have considered the processes in which collisions of the
Goldstone bosons on a metastable string induce its destruction by creating a
critical gap of the stable phase with lower tension. It has been shown that
at energy that is much smaller than the scale set by the tension $\e_1$ of the
string the interaction of the Goldstone bosons in the bulk of the string can be
neglected, and the discussed processes proceed `through the bounce', which
interaction mechanism has the scale set by the length $\ell_c$ of the critical
gap. We have demonstrated that when the energy scale $\omega$ in a collision is
much smaller than $1/\ell_c$ the probability $W_n$ of the string destruction in
an $n$ particle collision is proportional to $(\omega \ell_c)^{3n}$. In
particular the low-energy expression for $W_2$ is determined from our previous
result for the string decay at finite temperature. In order to calculate the
probability for an arbitrary relation between the energy scale and $\ell_c$ we
have extended the calculation of the thermal decay rate by formally introducing
a negative chemical potential for the bosons. The dependence of the resulting
string decay rate on the chemical potential and the temperature allows to
separate the contribution to this rate of individual $n$-particle collision
processes. In other words, the expression (\ref{dexp}) provides a generating
function for the probabilities $W_n$. It turns out that collisions of any odd
number of the Goldstone bosons do not lead at all to the string destruction and
this process occurs only in collisions of an even number of particles. As an
application of the general formulas we have calculated explicitly the
probability of destruction of the string in a two-particle collision as a
function of their c.m. energy $\sqrt{s}$. The found expression exhibits an
exponential growth at large values of the product $\sqrt{s} \, \ell_c$, however
within the validity of the approximations used in the present treatment this
growth never compensates the overall semiclassical exponential suppression of
the string decay.

\section*{Acknowledgments}
This work is supported  in part by the DOE grant DE-FG02-94ER40823. The work of
A.M. is also supported in part by the RFBR Grant No. 07-02-00878 and by the
Scientific School Grant No. NSh-3036.2008.2.

\end{document}